\documentclass[letterpaper,journal]{IEEEtran}
\usepackage[english]{babel}
\usepackage{microtype}
\usepackage{tikz}
\usepackage{amsmath,amsfonts,amsthm,mathtools}
\usepackage{bbm}
\usepackage{algorithmic}
\usepackage{algorithm}
\usepackage{array}
\usepackage[caption=false,font=normalsize]{subfig}
\usepackage{textcomp}
\usepackage{verbatim}
\usepackage{graphicx}
\usepackage{hyperref}

\hypersetup{%
  colorlinks=true,%
  linkcolor=red!50!black,
  citecolor=blue!65!black,
  urlcolor=blue!80!black,
  bookmarksnumbered=true,%
  bookmarksopen=true}

\newcommand\bs{\boldsymbol{s}} 
\newcommand\bu{\boldsymbol{u}} 
\newcommand\bp{\boldsymbol{p}} 
\newcommand\R{\mathbb{R}}
\newcommand\N{\mathbb{N}}
\newcommand\ip[2]{#1\cdot #2} 
\DeclarePairedDelimiter\prn()
\DeclarePairedDelimiter\bkt[]

\newtheorem{problem}{Problem}
\theoremstyle{definition}
\newtheorem{definition}{Definition} 

\begin{document}

\title{Scheduling the Charge of Temporally Flexible Electric \\ Vehicles: a Market-based Approach}

\author{Sabri El Amrani, Thibaut Horel, Saurabh Vaishampayan, Maryam Kamgarpour ~\IEEEmembership{Senior Member,~IEEE}, Munther A. Dahleh ~\IEEEmembership{Fellow,~IEEE}
        
\thanks{This work was supported by the Zeno Karl Schindler Foundation. \textit{(Corresponding author: Sabri El Amrani.)}}
\thanks{Sabri El Amrani, Saurabh Vaishampayan and Maryam Kamgarpour are with the Automatic Control Laboratory, EPFL, 1015 Lausanne, Switzerland (email: sabri.elamrani@epfl.ch, saurabh.vaishampayan@epfl.ch, maryam.kamgarpour@epfl.ch).}
\thanks{Thibaut Horel and Munther A. Dahleh are with the Laboratory for Information and Decision Systems, MIT, Cambridge, MA 02139 USA (email: thibauth@mit.edu, dahleh@mit.edu)}}

\maketitle

\begin{abstract}
The increasing electrification of human activities and the rapid integration of variable renewable energy sources strain the power grid. A solution to address the need for more grid storage is to use the battery of electric vehicles as a back-up capacity. However, drivers tend to disconnect their electric vehicle when its battery is needed the most. We propose a charge scheduler that incentivizes drivers to delay their disconnection to improve vehicle-to-grid services. We also leverage drivers' temporal flexibility to alleviate congestion in oversubscribed charging stations. We formulate the computation of an optimal flexible schedule as a mixed-integer quadratic problem. We tractably approximate its solution using the Alternating Direction Method of Multipliers. Considering the possibility that strategic drivers misreport their charging preferences to the station coordinator, we then propose a Vickrey–Clarke–Groves mechanism that incentivizes truthful reporting. We conclude with a simulated case study using real-world data to quantitatively assess the added value of drivers' temporal flexibility for enhancing vehicle-to-grid services and reducing station congestion.
\end{abstract}


\section*{Nomenclature}

\begin{table}[ht]
\centering
\textbf{Global Parameters} \\ [1ex]
\begin{tabular}{p{1.2cm} p{6.0cm}}
    $T$          & Number of discrete time intervals in a schedule \\
    $\delta t$   & Duration of one discrete time interval \\
    $N$          & Number of electric vehicles (EVs) at the station \\
    $\bp$        & Vector of day-ahead energy prices \\
    $C_{\mathrm{bus}}$ & Power capacity limit of the station's bus \\
    $\theta$     & Collection of the EVs' private types \\
    $x$          & Collection of the EVs' allocations \\
    $x^*(\cdot)$ & Socially efficient allocation \\
    $m$          & Collection of the EV payments \\
    $v_0(\cdot)$ & Utility of the EV charging station (EVCS) \\
\end{tabular}
\end{table}

\begin{table}[ht]
\centering
\textbf{Parameters of Electric Vehicle $n$} \\ [1ex]
\begin{tabular}{p{1.2cm} p{6.0cm}}
    $\eta_n$     & Power transfer efficiency \\
    $B_n$        & Battery capacity \\
    $s_n^0$      & Initial state of charge \\
    $b_n$        & Battery wear cost coefficient \\
    $s_n^d$      & Desired state of charge (SoC) at disconnection \\
    $\tau_n^d$   & Desired disconnection time \\
    $\alpha_n$   & Temporal inflexibility coefficient \\
    $\beta_n$    & SoC inflexibility coefficient \\
\end{tabular}
\end{table}

\begin{table}[ht]
\centering
\begin{tabular}{p{1.2cm} p{6.0cm}}
    $\theta_n$   & Private type \\
    $\tau_n$     & Actual assigned disconnection time \\
    $\bu_n$      & Power profile \\
    $x_n$        & Allocation \\
    $\bs_n$      & State of charge over time \\
    $r_n^c$      & Charging rate \\
    $r_n^d$      & Discharging rate \\
    $c_n(\cdot)$ & Cost function \\
    $v_n(\cdot)$ & Utility \\
    $m_n$        & Payment to the EVCS \\
\end{tabular}
\end{table}

\section{Introduction}
\IEEEPARstart{E}{lectric} vehicles (EVs) have become the fastest growing segment of the automobile industry, reaching double-digit market share in most countries \cite[p.~57]{pathak_climate_2022}. Uncoordinated charging of these EVs could strain the grid. For instance, locally optimized controls could increase peak net demand by 25\% in some locations according to projections for 2035 made in \cite{powell_charging_2022}.

If well managed, the generalized adoption of EVs offers an opportunity, however. Their deployment could indeed reduce the need for the costly large-scale electricity storage that the grid requires to integrate variable renewable energy sources. 
According to a recent estimate \cite{satchidanandan_efficient_2022}, the entire daily electricity demand of a state could be met by discharging 64\% of its cars, if they were all electric. The need for storage can hence be alleviated by using EVs' batteries as a back-up capacity for the grid. This vehicle-to-grid (V2G) service is referred to as \emph{arbitrage}. EVs are charged when energy prices are low, and discharged for a profit when they rise again.

The charge scheduling of EVs under grid constraints has been the subject of extensive literature, see \cite{mukherjee_review_2015} for a review. As an illustrative example, the work in \cite{ma_decentralized_2013} proposes a decentralized valley-filling algorithm that models EVs as energy-minimizing agents coupled through a common electricity price. In \cite{li_integrating_2024}, the scheduling problem is extended to bidirectional charging. The authors argue that successful V2G adoption requires fulfilling drivers' charge requests and compensating them for additional battery wear. Conversely, the work in \cite{liu_two-stage_2019} considers EVs with flexible charge requests. This flexibility is exploited in real-time by an aggregator to participate in the regulating power market.

These works have not considered the potential benefits of drivers' \textit{temporal flexibility}, however.
Drivers tend to pick up their car when the energy stored in their battery is needed the most (electricity demand typically peaks around 5--6 PM, when most workers leave the office) \cite{lee_acn-data_2019, iso_new_england_pricing_2024}.  To use EV batteries as a meaningful backup for the grid, a station coordinator should therefore be able to delay the disconnection of some EVs when needed. Delays could also alleviate congestion in charging stations when drivers are willing to postpone their disconnection to obtain a full charge. A recent survey \cite{suel_vehicle--grid_2024} suggests that EV drivers would be willing to offer this temporal flexibility in exchange for the right compensation.

This survey also shows that the requested compensation is different for each driver, it is a \textit{heterogeneous} preference parameter.
This cost of flexibility is also a \textit{private} parameter that is a priori unknown to the station coordinator. Without appropriate incentive scheme, strategic agents would benefit from reporting a flexibility that is lower than their real preference. This reduces their chance of being delayed by the coordinator. As a result, the successful deployment of temporally flexible charging schedules requires a mechanism for eliciting drivers' private preferences.

The EV scheduling literature has addressed how to exploit and elicit strategic drivers' desired disconnection time \cite{satchidanandan_efficient_2022} and the flexibility of their charge request \cite{de_hoog_market_2016, bhattacharya_extended_2016}.
Regarding drivers' \textit{temporal} flexibility, the authors of \cite{diaz-londono_mixed_2023} propose a framework where the disconnection of EVs is delayed to provide ancillary services (e.g.\ frequency regulation) to the grid. Drivers are incentivized to shift their disconnection time with a reward that is proportional to the duration of their delay. The interaction between the station coordinator and the EV drivers is modeled as a Stackelberg game in which the coordinator computes the coefficient of proportionality that maximizes its profit. This profit is also proportional to the duration of the EVs' delay. The EV drivers react by choosing the delay that maximizes their utility given their temporal flexibility parameter. However, to be able to compute the optimal reward coefficient, the coordinator is assumed to have knowledge of the drivers' flexibility parameters. This assumption ignores the possibility of untruthful reporting of this private preference. Moreover, assuming that the station's profit is proportional to the duration of the EVs' delay might be reasonable for ancillary services like frequency regulation, but it is an inadequate model of revenues from arbitrage.

In \cite{van_huffelen_grid-constrained_2025} and \cite{tsaousoglou_max-min_2020}, EVs are delayed if grid constraints prevent the full charge of all their batteries in time. In these works, the drivers are implicitly assumed to have the same flexibility. The disconnection of each EV is a variable controlled by the station coordinator to minimize the sum of all delays \cite{van_huffelen_grid-constrained_2025}, and to minimize the maximum delay \cite{tsaousoglou_max-min_2020}. However, if the assumption of homogeneous flexibilities fails to hold, either of these schemes might result in the delay of urgent charge requests instead of more flexible ones. 

The work in \cite{lv_optimal_2024} approaches the problem differently. The coordinator uses price signals to guide the behavior of a fleet of EVs towards charging either on their way to work or on their way back home. Instead of a delay, temporal flexibility thus takes the form of a binary decision. The station coordinator optimizes prices as a function of grid congestion levels to maximize its profit. Rather than individual preferences, the coordinator is assumed to have access to a model of the aggregate behavior of the fleet as a function of prices. The main downside of this coarser approach is that it makes energy arbitrage impossible. The coordinator needs to know individual disconnection times to be able to overcharge the EVs that can be used as back-up for the grid.

In summary, the following gaps seem to emerge in the literature on the scheduling of temporally flexible EVs. First, there appears to be no study on the benefits of delaying EVs to boost arbitrage schemes. Second, existing work assumes that drivers always require a full charge and that they want to disconnect their EV as soon as it is charged. It does not address the trade-off between temporal an charge flexibility, and it ignores the challenge of eliciting drivers' desired disconnection time. Finally, the elicitation of heterogeneous drivers' temporal flexibility has yet to be addressed.

We propose a model for the scheduling of EVs with both flexible disconnection times and flexible charge requests. We also address the joint elicitation of all the drivers' private charging preferences, including their temporal flexibility and their desired disconnection time. Our framework is applicable to the provision of V2G services and to the alleviation of station congestion.
Our main contributions are the following.
\begin{itemize}
    \item Formulation of a bidirectional charge scheduling problem for a shared, capacity-limited EV charging station. The formulation models the EVs' battery wear as well as the drivers' flexible disconnection time and SoC requests,
    \item Empirical assessment of the practical value of EV drivers' temporal flexibility for arbitrage and station congestion relief, supported by simulations with real-world data,
    \item Application and empirical study of a mechanism that truthfully elicits the EVs' private charging preferences while ensuring voluntary participation.
\end{itemize}

The paper is structured as follows. Section~\ref{sec:scheduling} formulates the flexible bidirectional charge scheduling problem, and proposes an efficient method for approximating its solution. Section~\ref{sec:mechanism} describes the preference elicitation problem and presents the mechanism that incentivizes truthful reporting. Section~\ref{sec:case_study} is a simulated case study which assesses the potential benefits of a real-world deployment of V2G with flexible schedules.

\section{Scheduling the Charge of Flexible Electric Vehicles}
In this section, we start by presenting our model of a capacity-limited charging station hosting a fleet of flexible EVs. We then formulate the bidirectional flexible scheduling problem. We conclude by proposing a heuristic for approximating its solution.

\label{sec:scheduling}
\subsection{The Scheduling Problem}
\subsubsection{The Models of the Charging station and the Electric Vehicles}
We consider $N\in\N$ electric vehicles connecting to the same charging station (EVCS), whose coordinator centrally schedules the vehicles' charging over the course of a day. The day is split in $T \in \N$ intervals of equal duration $\delta t\in\R_+$ and indexed from $0$ to $T-1$. We assume that the vector of energy prices in each time interval $\bp \in \R^T$ was fixed in the day-ahead market and is known to the EVCS coordinator when the day starts. The EVCS is assumed to have one power bus of capacity $C_{\mathrm{bus}} \in \R^+$ (charges are expressed as power flows throughout this work).

\begin{figure}[H]
\centering
\begin{tikzpicture}[scale=0.487]
    \definecolor{connected}{RGB}{144, 238, 144}
    \definecolor{disconnected}{RGB}{255, 99, 71}
    \def\N{5}  
    \def\T{7}  

    \foreach \n in {1,3,5} {
        \foreach \t in {1,2,3} {
            \fill[connected] (\t,-\n) rectangle ++(0.8,0.8);
        }
        \foreach \t in {6,7} {
            \fill[disconnected] (\t,-\n) rectangle ++(0.8,0.8);
        }
    }
    \fill[connected] (5,-1) rectangle ++(0.8,0.8);
    \fill[connected] (6,-1) rectangle ++(0.8,0.8);
    \fill[disconnected] (7,-1) rectangle ++(0.8,0.8);
    \fill[connected] (5,-3) rectangle ++(0.8,0.8);
    \fill[connected] (6,-3) rectangle ++(0.8,0.8);
    \fill[connected] (7,-3) rectangle ++(0.8,0.8);
    \fill[connected] (5,-5) rectangle ++(0.8,0.8);
    \fill[disconnected] (6,-5) rectangle ++(0.8,0.8);
    \fill[disconnected] (7,-5) rectangle ++(0.8,0.8);  

    \node[left] at (2.9,-1.6) {...};
    \node[left] at (2.9,-3.6) {...};
    \node[left] at (6.9,-1.6) {...};
    \node[left] at (6.9,-3.6) {...};
    \node[left] at (4.9,-4.6) {...};
    \node[left] at (4.9,-2.6) {...};
    \node[left] at (4.9,-0.6) {...};

    \node[left] at (3.2,-6.0) {...};
    \node[left] at (6.2,-6.0) {...};

    \node[left] at (0.5,-0.6) {1};
    \node[left] at (0.6,-1.6) {...};
    \node[left] at (0.5,-2.6) {$n$};
    \node[left] at (0.6,-3.6) {...};
    \node[left] at (0.6,-4.6) {$N$};
    \node[left] at (-0.7,-2.6) {EVs};

    \foreach \t in {1,4,7} {
        \node[below] at (\t+0.4,-5.5) {\ifnum\t=1 0\else\ifnum\t=4 $t$\else {\small $T-1$}\fi\fi};
    }
    \node[left] at (10,-5.15) {\small Time};

    \fill[connected] (10,-2.5) rectangle ++(0.8,0.8);
    \node[right] at (10.8,-2.1) {\small EV connected};
    \fill[disconnected] (10,-3.5) rectangle ++(0.8,0.8);
    \node[right] at (10.8,-3.1) {\small EV disconnected};

    \draw[thick,->] (1,-5.2) -- (8,-5.2);

\end{tikzpicture}
\caption{A schematic representation of the model of the charging station.}
\label{fig:scheduling_diagram}
\end{figure}
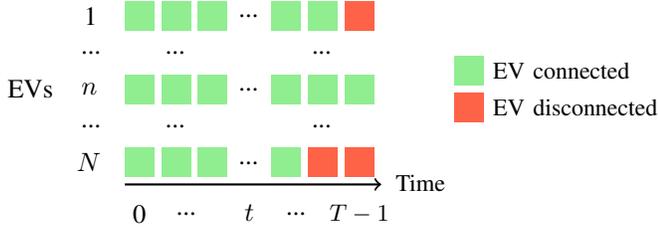

For simplicity, we assume that all the EVs simultaneously connect to the station, that their disconnection is final, and that their charging spot is not taken up by another car after their departure. An EV's connection schedule is therefore entirely described by one parameter: its disconnection time. A similar model was first proposed by \cite{satchidanandan_efficient_2022}. It is illustrated in Figure~\ref{fig:scheduling_diagram}.

Each EV $n \in \{1, \dots, N\}$ is characterized by some public parameters (known to the coordinator) and private preferences (unknown to the coordinator). The public parameters of EV $n$ are its battery capacity $B_n \in \R^+$, battery wear coefficient $b_n \in \R^+$, initial SoC, $\bs_n[0] \in [0, B_n]$, power transfer efficiency $\eta_n \in [0,1]$, maximum charging rate $r_n^c \in \mathbb{R}^+$ and maximum discharging rate $r_n^d \in \mathbb{R}^+$.
The private preferences of EV $n$, also referred to as its \emph{type}, includes the desired disconnection time $\tau^d_n \in \{0, \dots, T\}$, desired SoC at disconnection $s^d_n \in [0, B_n]$, temporal inflexibility $\alpha_n \in \R^+$ and SoC inflexibility $\beta_n \in \R^+$. We denote the type of EV $n$ as $\theta_n = (\tau^d_n, s^d_n, \alpha_n, \beta_n) \in \Theta_n$ and the vector of all types as $\theta = (\theta_1, \dots, \theta_N) \in \Theta$.

\subsubsection{The Scheduling Decision}
\looseness=-1
The goal of the coordinator is to choose, for each EV $n$, its actual disconnection time $\tau_n \in \{0, \dots, T\}$ and power schedule $\bu_n \in \R^T$. The decision $x_n = (\tau_n, \mathbf{u}_n) \in X_n$ is referred to as the \textit{allocation} to EV $n$ and $x = (x_1, \dots, x_n) \in X$ is the vector of all allocations.
We assume throughout this work that the EVs follow their allocation.\footnote{In practice, this could be enforced through various mechanisms (fines, threats of a temporary ban, etc.) that we do not explicitly model here. Alternatively, one could have adopted an information design approach in which the allocation is an action recommendation specifically designed to induce \emph{obedience}: conditioned on the available information, each EV's best-response is to follow their recommendation.}

For a given power schedule $\bu_n$, the SoC evolution of EV $n$ is given by the following equation:
\begin{equation}
    \bs_n[t] = \bs_n[0] + \eta_n \sum^{t-1}_{k=0} \bu_n[k] \ \delta t, 
    \label{eq:soc_evolution}
\end{equation}
for each $t \in \{0, \dots, T\}$ and where $\eta_n \in [0, 1]$ is the EV's power transfer efficiency. Moreover, the allocations must satisfy the following constraints: 
\begin{subequations}
\label{eq:constraints}
\begin{align}
&- r^d_n \leq \bu_n[t] \leq r^c_n, \qquad &&\forall n,\ \forall t, \label{eq:constraint_charge_rates} \\
&0 \leq \bs_n[t] \leq B_n, \qquad &&\forall n,\ \forall t, \label{eq:constraint_battery_capacity} \\
&\bu_n[t]\,\mathbbm{1}_{\{t \geq \tau_n\}} = 0, \qquad &&\forall n,\ \forall t, \label{eq:constraint_exchange} \\
    &\left| \sum_{n=1}^{N} \bu_n[t] \right| \leq C_{\mathrm{bus}}, \qquad &&\forall t.\label{eq:constraint_bus}
\end{align}
\end{subequations}
Equation \eqref{eq:constraint_charge_rates} ensures that the power charged to and from an EV does not exceed its charging and discharging rates $r^c_n, r^d_n \in \R^+$, respectively. Equation \eqref{eq:constraint_battery_capacity} enforces each battery's capacity constraint. Equation \eqref{eq:constraint_exchange} ensures that no power exchange occurs after an EV's disconnection. Equation \eqref{eq:constraint_bus} enforces the capacity of the station's bus.

\subsubsection{The Scheduling Objective} 
The utility of EV $n$ for its allocation is described by the cost function $c_n: X_n \times \Theta_n \to \R$ that it wishes to minimize:
\begin{multline}
    \hspace{-0.7em}
    c_n(x_n, \theta_n) = \alpha_n \left(\tau_n - \tau_n^d\right)^2 + \beta_n \left(\max\{0, s^d_n - \bs_n[T]\}\right)^2 \\
    + b_n \|\bu_n\|_1 \ \delta t,
    \label{eq:ev_cost}
\end{multline}
where $\| \cdot \|_1$ is the $\ell_1$-norm.
The first two terms express the driver's dissatisfaction for not disconnecting at the desired time and for an incomplete charge, respectively. The terms are quadratic to model increasing marginal dissatisfaction as is commonly assumed in preference models.
Battery wear is modeled as an $\ell_1$-norm and is thus proportional to the amount of transferred energy. This simplifying model notably ignores the effect of deep discharges on the battery's life \cite{visakh_profitability_2022}.

We assume in this work that the EVCS coordinator is benevolent. Instead of maximizing its profit, it seeks to maximize \emph{social welfare}, or equivalently, minimize the total cost, that is, the sum of the energy cost, namely $\sum_{n=1}^N \ip\bp\bu_n$, and the costs of the $N$ EVs.
This choice of objective thus results in the following formulation of the scheduling problem.

\vspace{0.5em}

\begin{problem}[Flexible scheduling]
\label{pb:flexible_scheduling}
\begin{equation} 
    \min_{x \in X} \sum_{n=1}^N \left[c_n(x_n, \theta_n) + \ip\bp\bu_n\right],
    \label{eq:social_welfare}
\end{equation}
subject to constraints \eqref{eq:constraint_charge_rates} to \eqref{eq:constraint_bus}.
\end{problem}

\vspace{0.5em}

An allocation that maximizes social welfare, that is, a solution to Problem~\ref{pb:flexible_scheduling}, is called \textit{socially efficient} and we write it as $x^*(\theta)$. Note that this solution depends on the EVs' private types $\theta$ which are a priori unknown to the EVCS coordinator. In the following section, we relax this assumption and present an algorithm that approximately computes the efficient allocation assuming knowledge of $\theta$. In Section~\ref{sec:mechanism}, we return to the general case and present a mechanism that allows the EVCS to truthfully elicit the private types $\theta$.

\subsection{A Decomposition Technique for Efficient Scheduling}
The cost functions of the EVs \eqref{eq:ev_cost} include a term that is quadratic in the disconnection time, which is a discrete variable. Consequently, Problem \ref{pb:flexible_scheduling} is a mixed-integer quadratic problem, which is generically NP-hard \cite{pia_mixed-integer_2014}. 
In the worst case, an exact method like Branch-and-Bound would need to solve a quadratic program for each of the $(T+1)^N$ possible combinations of disconnection time values to find the optimal solution.

Instead, we propose using the Alternating Direction Method of Multipliers (ADMM) as an approximation heuristic. This iterative Lagrangian method decouples the schedules of the $N$ individual EVs by relaxing the power capacity constraint of the EVCS' bus \eqref{eq:constraint_bus}. The algorithm alternates between solving decoupled scheduling problems, one for each EV, and a dual ascent step that penalizes violations of the coupling constraint.

Writing $z_+=\max\{0,z\}$ for the positive part of the scalar $z$, the subproblem of EV $n\in\{1,\dots,N\}$ at iteration $k+1$ consists in minimizing the augmented cost
\begin{multline}
    c_n(x_n, \theta_n) + \ip\bp\bu_n  - \frac 1{2\nu}\sum_{t=0}^{T-1} (\lambda_t^k)^2
	\\
	+ \frac 1{2\nu}\sum_{t=0}^{T-1}\Bigl[\lambda_t^k + \nu\Bigl(\Bigl|\bu_n[t] + \sum_{\substack{j=1 \\ j \ne n}}^N \hat\bu_j[t]\Bigr| 
    - C_{\mathrm{bus}}\Bigr)\Bigr]_+^2
\label{eq:admm_subproblem}
\end{multline}
over the variable $x_n =(\bu_n,\tau_n)$ and subject to the constraints \eqref{eq:constraint_charge_rates} to \eqref{eq:constraint_exchange}. Here, $\nu \in \R$ is a penalty parameter, $\lambda^k_t \in \R$ are dual variables and
\[
\hat{\bu}_j[t] =
\begin{cases}
\bu_j^{k+1}[t] & \text{if } j < n \\
\bu_j^{k}[t]   & \text{if } j > n
\end{cases}.
\]
Denoting by $x_n^{k+1}$ the solution of the above problem, the dual ascent step at iteration $k+1$ is given by
\begin{equation}
    \lambda_t^{k+1} = \Big[\lambda_t^k + \nu \Big(\Big| \sum\nolimits_{n=1}^{N} \bu_n^{k+1}[t] \Big| - C_{\mathrm{bus}} \Big) \Big]_+,
\label{eq:admm_dual_update}
\end{equation}
with initialization $\lambda^0_t = 0$ for $t\in\{1,\dots,T\}$.

These two steps repeat until a stopping criterion is met. While ADMM is not guaranteed to converge to an optimal solution for non-convex problems (including mixed-integer problems), it is commonly used as an approximate heuristic \cite{maneesha_survey_2021}.
For our scheduling problem, empirical comparisons suggest that the cost achieved by ADMM's approximate solution is, on average, only 2.4\% higher than the optimum. This makes ADMM a viable heuristic for approximately solving Problem \ref{pb:flexible_scheduling}.

\section{Eliciting Electric Vehicle Drivers' Private Charging Preferences}
\label{sec:mechanism}

As noted below Problem~\ref{pb:flexible_scheduling}, computing the socially efficient allocation requires knowledge of the EVs' private types $\theta$. In the present section we adopt a mechanism design approach and design payments from the EVs to the station that result in the truthful elicitation of the private types and implement the socially efficient outcome.

\subsection{The Mechanism Design Problem}

While a general mechanism allows arbitrary communication between the drivers and the charging station, we restrict ourselves to \emph{direct mechanisms} in which the drivers communicate their preference reports $\hat{\theta}_n = (\hat\tau^d_n, \hat s^d_n, \hat\alpha_n, \hat\beta_n) \in \Theta_n$ to the coordinator when connecting to the station.
We denote the vector of reports as $\hat{\theta} = (\hat{\theta}_1, \dots, \hat{\theta}_N) \in \Theta$. Using these reports, the coordinator computes, for each EV $n$, its allocation $\tilde x_n(\hat\theta)$ and collects a payment $m_n(\hat\theta_n)$. This is summarized in the following definition.

\begin{definition}[Mechanism]
    \label{def:mechanism}
    A (direct) mechanism is a collection of $n$ \textit{allocation rules} $\tilde x_n: \Theta \rightarrow X$ and $n$ \textit{payment rules} $m_n: \Theta \rightarrow \R$.
\end{definition}

By the Revelation Principle \cite[\S 7.2]{fudenberg_game_1991}, the restriction to direct mechanisms is without loss of generality. This is because, for any mechanism, there exists a direct mechanism that results in the same outcome. Such a direct mechanism must however account for the possibility of strategic misreporting from the drivers by satisfying the following two properties:
\begin{enumerate}
    \item \textit{Dominant-strategy incentive-compatibility (DSIC).} Reporting its true preference should maximize an EV's utility, regardless of the reports of the other EVs.
    \item \textit{Individual rationality}. An EV should always have a greater utility in participating in the mechanism than remaining discharged.\footnote{This constraint can be adapted if an EV has a broader range of options than either connecting to the present EVCS or remaining discharged.}
\end{enumerate}

We assume that the allocations $\tilde x_n(\hat\theta) = (\tilde \tau_n(\hat\theta), \tilde \bu_n(\hat\theta))$ and the payments $m_n(\hat\theta)$ enter the utility $v_n: X_n \times \Theta_n \times \R \to \R$ of EV $n$ in the following manner:
\begin{equation}
    v_n (\tilde x_n(\hat\theta), \theta_n, m_n(\hat\theta)) = -c_n( \tilde x_n(\hat\theta), \theta_n) - m_n(\hat\theta).
    \label{eq:ev_utility}
\end{equation}
Utilities of this shape are called \emph{quasi-linear} in the game theory literature \cite{li_transactive_2020}, as they are linear in the payments.

Denoting by $\tilde x$ the vector of allocation rules and $m$ the vector of payment rules, the mechanism design problem takes the following form.


\vspace{0.5em}
\begin{problem}[Mechanism design]
    \label{pb:mechanism_design}
    \begin{equation}
        \label{eq:social_welfare_mechanism}
        \min_{\substack{\tilde x, m}} \sum_{n=1}^N \left[c_n(\tilde x_n(\hat \theta), \theta_n) + \ip\bp\tilde\bu_n(\hat\theta)\right],
    \end{equation}
    subject to constraints \eqref{eq:constraint_charge_rates} to \eqref{eq:constraint_bus}, and for each  $1\leq n\leq N$ and $\hat\theta_n\in\Theta_n$,
    \begin{align}
        &v_n\big(\tilde x_n(\theta), \theta_n, m_n(\theta)\big) \geq v_n\big(\tilde x_n(\hat\theta), \theta_n, m_n(\hat\theta)\big),\label{eq:ic}\\
        &v_n\big(\tilde x_n(\theta), \theta_n, m_n(\theta)\big) \geq -\beta_n \big(s^d_n - \bs_n[0]\big)^2. \label{eq:ir}
    \end{align}
\end{problem}
\vspace{0.5em}

The mechanism design problem is similar to Problem~\ref{pb:flexible_scheduling} with two additional constraints guaranteeing dominant-strategy incentive-compatibility \eqref{eq:ic} and individual rationality \eqref{eq:ir}. Another difference is that the optimization variables are now \emph{functions} mapping type reports to allocations and payments. Finally, observe that the payments do not appear the objective function: this is because they are counted negatively in the EVs' utilities but positively in the station utility and thus cancel out in the social welfare.

\subsection{The Vickrey–Clarke–Groves Mechanism}
\label{subsec:vcg}

The design of incentive-compatible mechanisms has been extensively studied in economics (see e.g.\ \cite[chap.~7]{fudenberg_game_1991}). We propose to solve Problem \ref{pb:mechanism_design} using the \textit{Vickrey--Clarke--Groves (VCG)} mechanism. For quasi-linear utilities, the VCG mechanism is known to be DSIC, elicits multidimensional types and attains the socially efficient outcome when the coordinator seeks to maximize social welfare \cite[\S 7.4.3]{fudenberg_game_1991}.

\begin{definition}[Vickrey--Clarke--Groves mechanism]
    \label{def:vcg}
    The allocation rule of the VCG mechanism maximizes social welfare assuming that preferences reports $\hat\theta$ are truthful:
    \begin{equation} 
        \tilde x(\hat\theta) = \min_{x \in X} \sum_{n=1}^N \bkt*{c_n(x_n, \hat\theta_n) + \ip\bp\bu_n},
        \label{eq:vcg_allocation} 
    \end{equation}
    under constraints \eqref{eq:constraint_charge_rates} to \eqref{eq:constraint_bus}, but ignoring constraints \eqref{eq:ic} and \eqref{eq:ir}. In other words, $\tilde x(\hat\theta)=x^*(\hat\theta)$.

    The payments are given by the following formula:
    \begin{multline}
        m_n(\hat\theta) = \ip\bp\tilde\bu_n(\hat\theta)
        +\sum_{\substack{j=1 \\ j \ne n}}^{N} \bkt*{c_j\big(\tilde x_j(\hat\theta), \hat\theta_j\big)
        -c_j\big( x_j^{(-n)}, \hat\theta_j \big)}\\[-1em]
        +\sum_{\substack{j=1 \\ j \ne n}}^N \ip\bp\prn*{\tilde \bu_j(\hat\theta) -\bu_j^{(-n)}},
        \label{eq:vcg_payment}
    \end{multline}
    where the superscript $^{(-n)}$ denotes a socially efficient allocation computed in the absence of EV $n$ (assuming truthful reports).
\end{definition}
 The VCG payment has a natural interpretation: each EV $n$ pays for the \emph{externalities} it imposes on the EVCS and the other EVs by being present. These include the cost of the energy used to charge EV $n$ (externality imposed on the station) as well as the changes in utilities and charging costs of the remaining EVs induced by $n$'s presence. These changes can be either positive or negative depending on whether $n$'s presence benefits the station (e.g.\ when using EV $n$'s battery to charge other EVs) or hampers it (e.g.\ by increasing congestion).

 As mentioned above Definition~\ref{def:vcg}, the VCG mechanism has the properties we desired:
\begin{enumerate}
    \item \textit{Dominant-strategy incentive-compatibility.} This follows from the quasi-linear utilities \eqref{eq:ev_utility} and the coordinator's objective of maximizing social welfare \eqref{eq:social_welfare_mechanism},
    \item \textit{Social efficiency.} As the VCG mechanism is DSIC, $\hat\theta = \theta$. Consequently, the VCG allocation \eqref{eq:vcg_allocation} is socially efficient: $\tilde x(\hat\theta) = \tilde x(\theta) = x^*(\theta)$.
    \item \textit{Individual rationality.}  We prove this below.
\end{enumerate}
    \begin{proof}
        \label{proof:ir}
        Substituting the expression of the VCG payment \eqref{eq:vcg_payment} into the individual rationality constraint \eqref{eq:ir}, we obtain:
        \begin{multline*}
        \sum_{j=1}^{N} \Bigl[ - c_j\bigl( x_j^*, \hat\theta_j \bigr)- \ip\bp\bu_j^*\Bigr]
                \geq - \beta_n\big(s^d_n - \bs_n[0]\big)^2\\
    	    - \sum_{\substack{j=1 \\ j \ne n}}^N \Bigl[
                c_j\bigl( x_j^{(-n)}, \hat\theta_j \bigr)
    	    + \ip\bp\bu_j^{(-n)} \Bigr].
        \end{multline*}
        The right-hand side of this inequality maximizes social welfare under the additional constraint that EV $n$ does not receive any charge. As this schedule belongs to the set $X$ of feasible schedules in Problem~\ref{pb:flexible_scheduling}, the less constrained optimal social welfare of the left-hand side is necessarily larger. Hence, the VCG payment is individually rational.
    \end{proof}
The VCG mechanism does not satisfy (weak) budget balance, however. This means that the cost of the energy used to charge the EVs might not be covered by the collected payments, resulting in a negative utility for the station \cite{li_transactive_2020}.

\begin{figure*}[!t]
\centering
\subfloat[]{
    \includegraphics[width=3in]{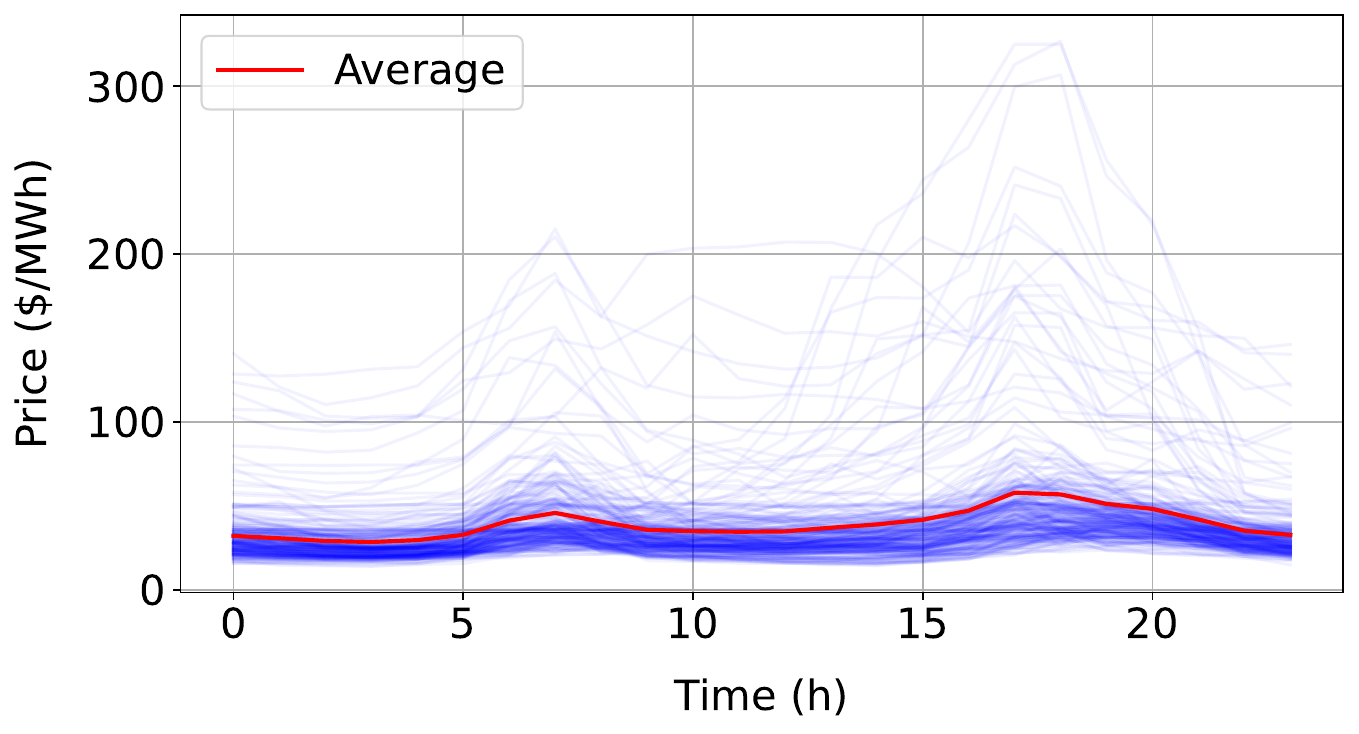}
}
\subfloat[]{
    \includegraphics[width=3in]{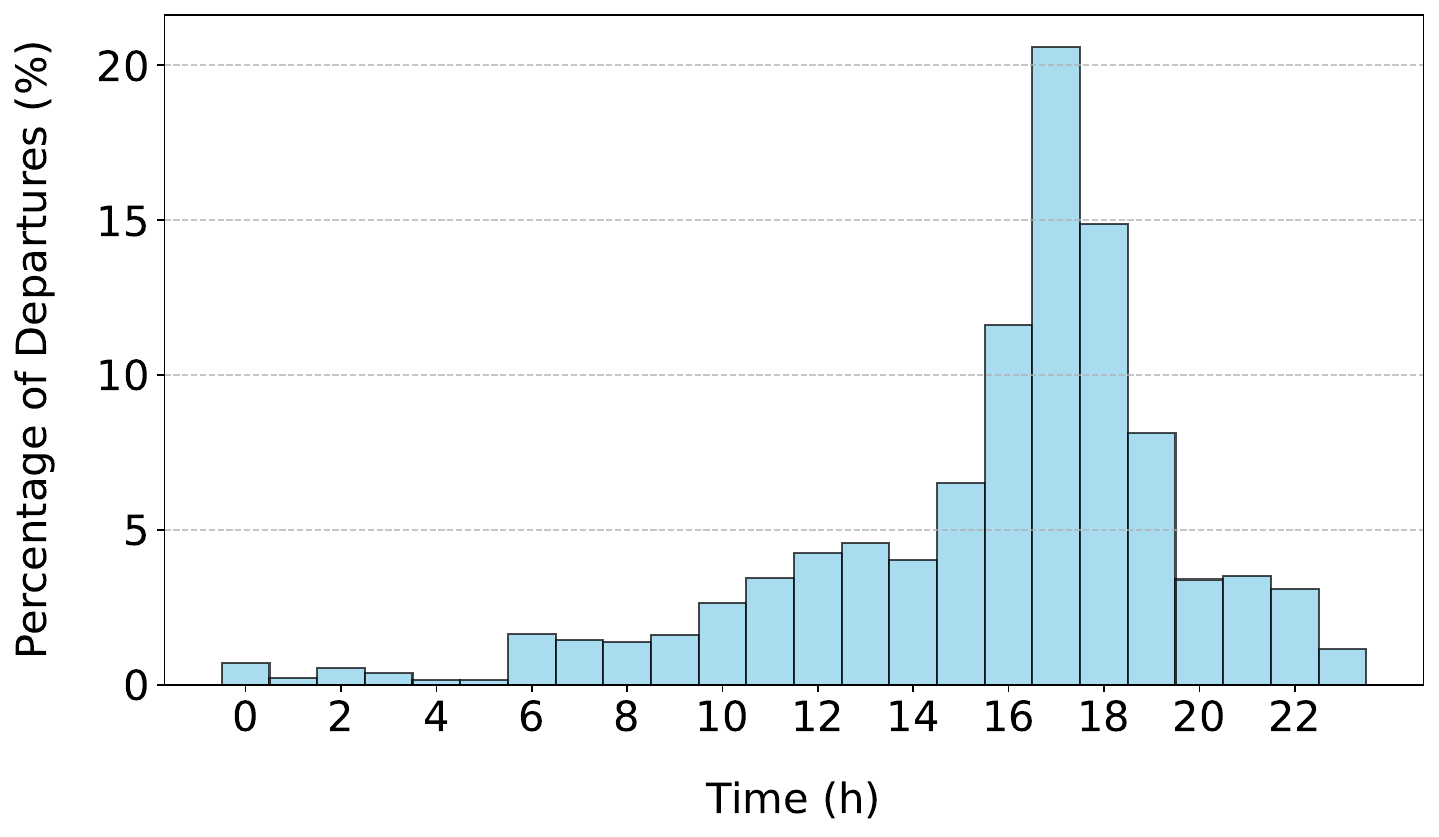}
}

\subfloat[]{
    \includegraphics[width=3in]{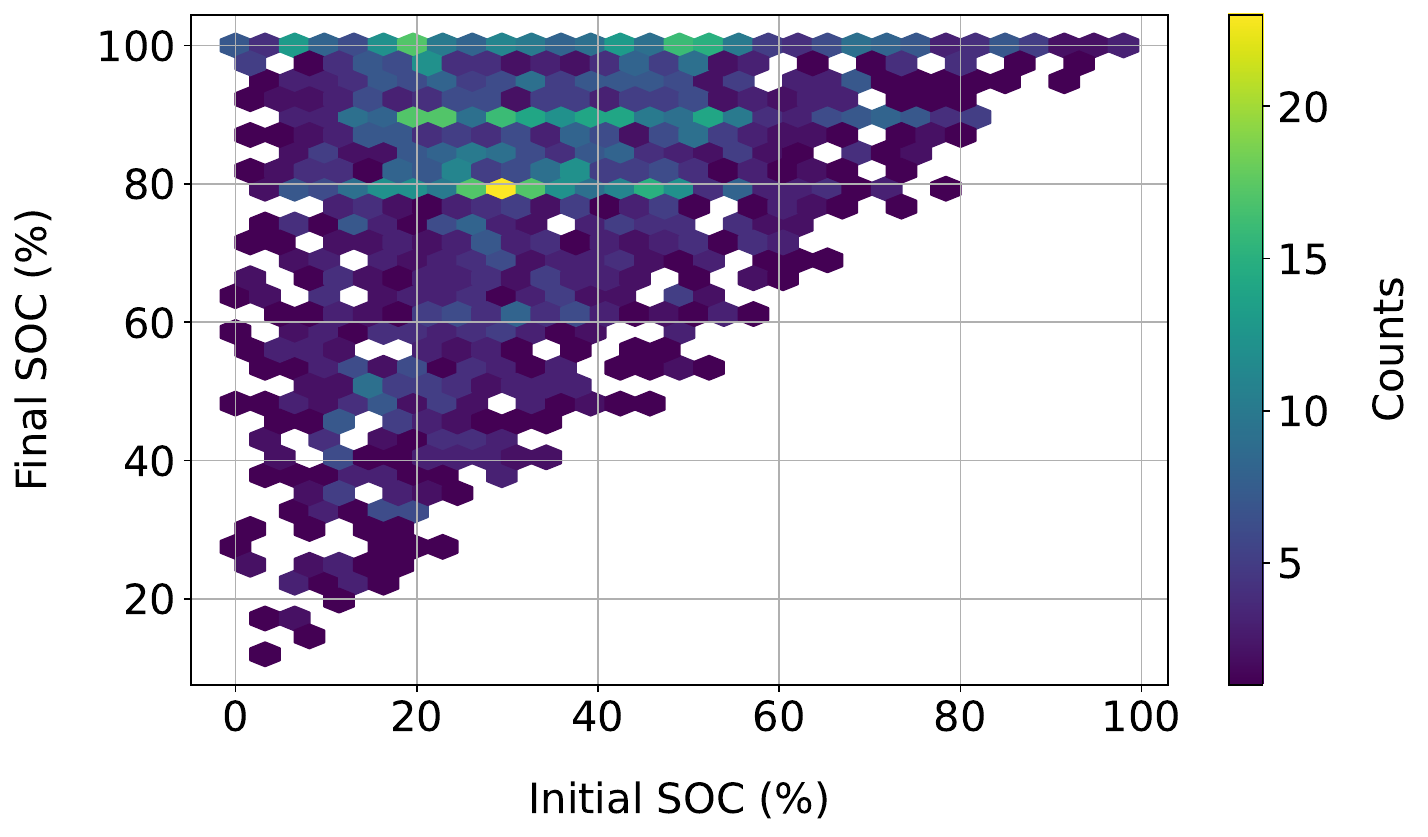}
}
\subfloat[]{
    \includegraphics[width=3in]{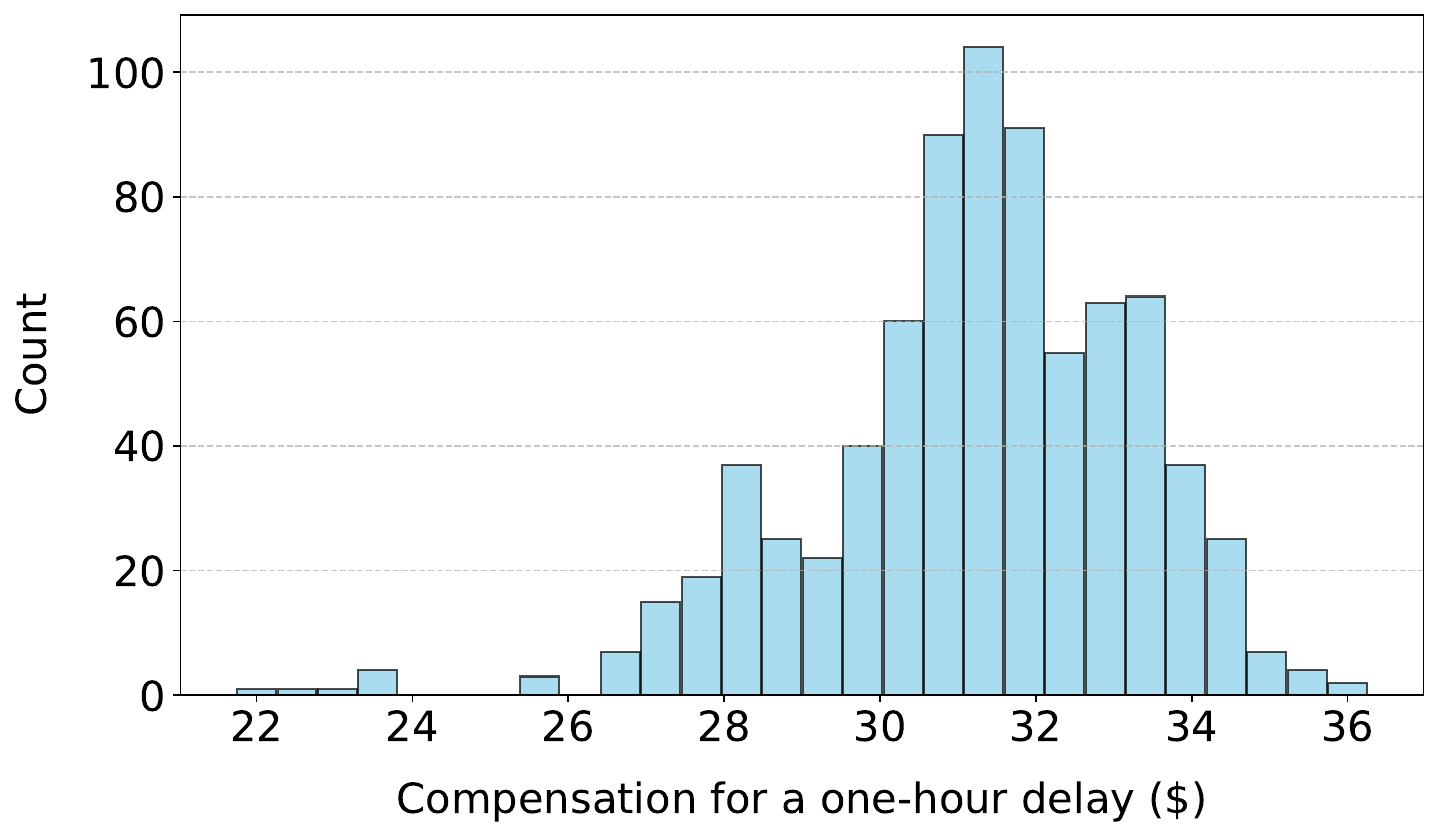}
}
    \caption{Real-world datasets used in our case study. (a) Market prices in New England (Nov.\ 1, 2023 to Oct.\ 31, 2024) \cite{iso_new_england_pricing_2024}. (b) EV disconnection times at a parking lot (Nov.\ 1, 2018 to Jan.\ 2, 2019)\cite{lee_acn-data_2019}. (c) Initial and final charge of EVs in 3 charging stations \cite{distributed_electrical_systems_laboratory_desl_epfl_desl-epfllevel-3-ev-charging-dataset_2025}. (d) Compensation requested by drivers to delay their scheduled departure by one hour according to a survey \cite{suel_vehicle--grid_2024}. This compensation for a one-hour delay also corresponds to the value of the inflexibility coefficient $\alpha_n$ in our EV cost model \eqref{eq:ev_cost}.}
\label{fig:data}
\end{figure*}

\section{Case Study}
In this section, we use simulations to assess the value of drivers' temporal flexibility in our bidirectional charging schedule. We start by describing the setup with an overview of the real-world data used in the experiments. We then study the effect of temporal flexibility on V2G profits and station congestion relief. Subsequently, we conduct simulations showing the need for truthful mechanisms and conclude with an empirical study demonstrating the benefits of the VCG tax.

\label{sec:case_study}
\subsection{Simulation Set-up}

Our numerical experiments are conducted in the following environment.\footnote{GitHub: \url{https://github.com/Sabri2001/v2g\_mechanism\_project}.} We consider a charging station to which 5 Nissan Leaf EVs simultaneously connect at 10 \textsc{a.m.} The day is assumed to end at 10 \textsc{p.m.}, and the experiments are run with a time granularity $\delta t$ of 15 minutes. The Nissan Leaf has a battery capacity $B_n$ of 40\,kWh, a power transfer efficiency $\eta_n$ of 0.87 and we estimated its battery wear cost coefficient $b_n$ to be of 0.13\,\$/kWh. This estimate was obtained by spreading the cost of replacing the battery over the car's expected lifetime mileage \cite{nissan_2026_2025}. The EVs are assumed to all have maximum charging and discharging rates (respectively $r^c_n$ and $r^d_n$) of 6.6\,kW, similarly to Lee et al.\ \cite{lee_acn-data_2019}. We mainly consider two scenarios for the power limit of the station's bus. In the \textit{regularly congested} setting, the station is assumed to have a power limit of 15\,kW. This corresponds to an oversubscription ratio of 2.2, as in the charging station in Lee et al.\ \cite{lee_acn-data_2019}. The oversubscription is the ratio between the number of EVs at the station and the maximum number of EVs that can be charged at full rate in parallel, given the capacity of the station's power bus $C_{bus}$. In the \textit{highly congested} setting, this limit is reduced to 10\,kW (which corresponds to an oversubscription ratio of 3.3). 

\begin{figure*}[!t]
\centering
    \subfloat[]{
    \includegraphics[width=3in]{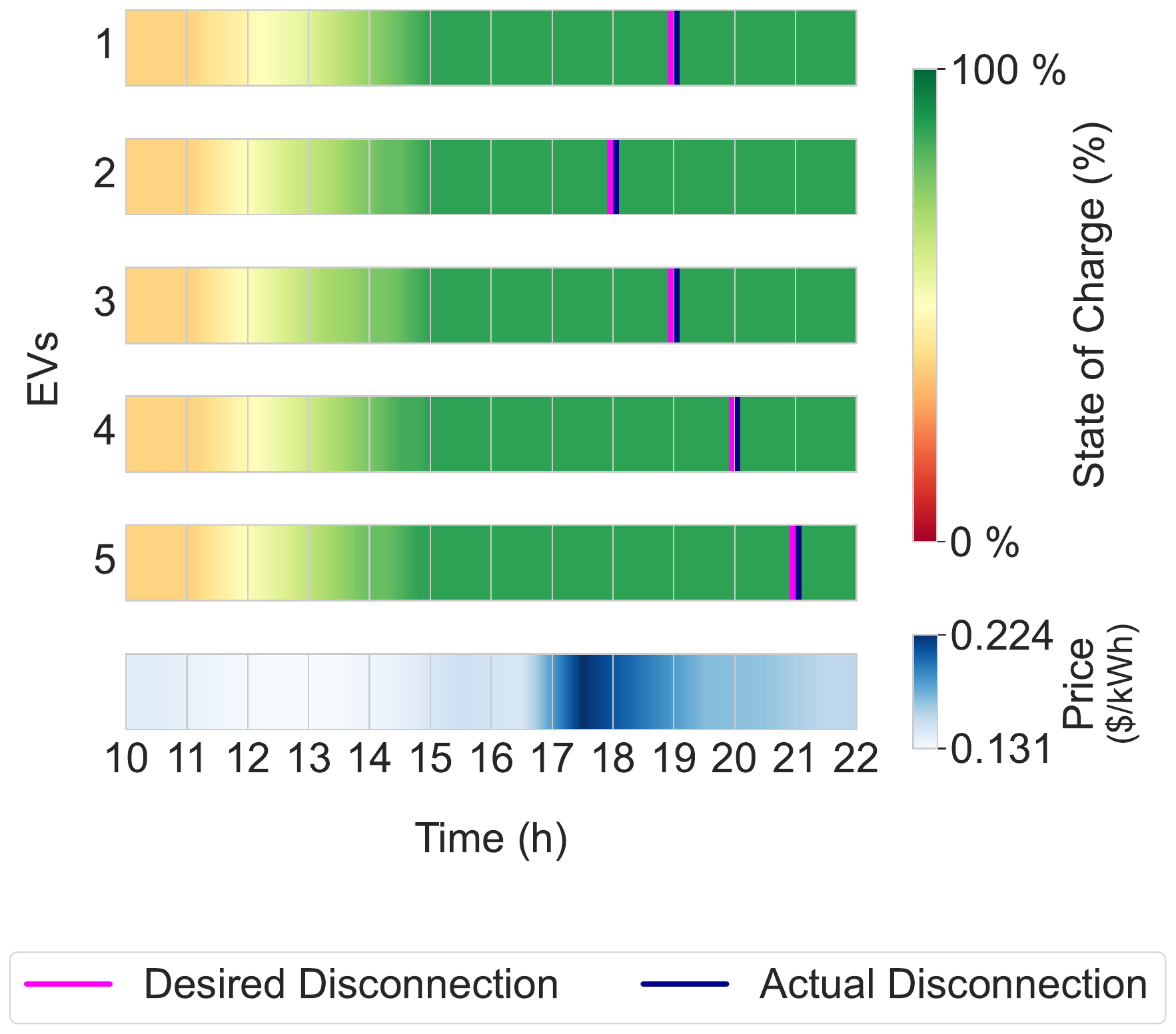}
}
\subfloat[]{
    \includegraphics[width=3in]{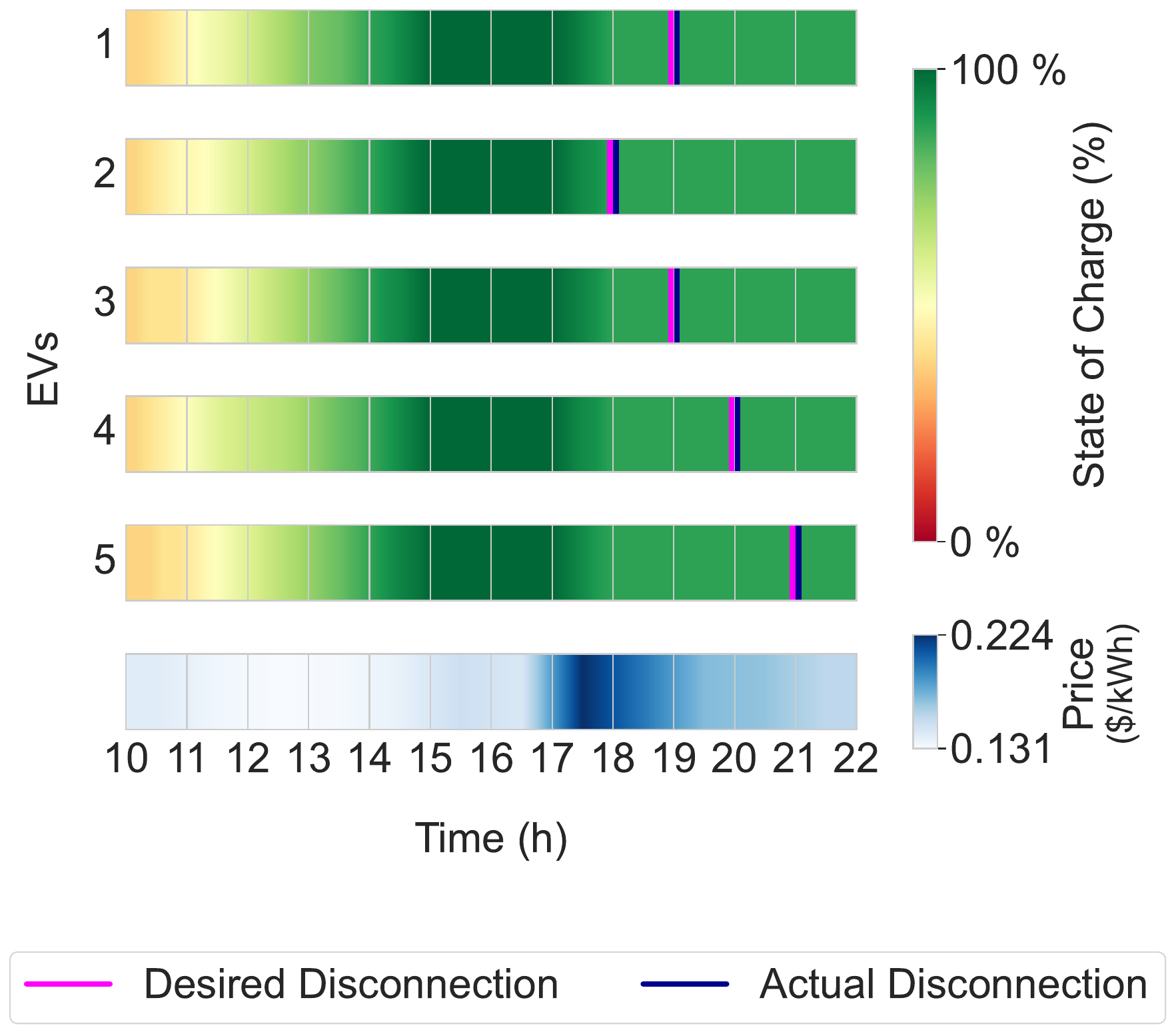}
}
\caption{Optimal flexible scheduling with energy market prices of Jan.\ 17, 2024, in a regularly congested station. (a) Bidirectional charging is not profitable with the real battery wear cost of 0.13\,\$/kWh. (b) Bidirectional charging becomes profitable when battery wear cost is reduced by 75\%.}
\label{fig:soc_battery}
\end{figure*}

Most of the experiments in this section compute average metrics over 20 runs with \textit{randomly sampled} data. We hereby mean that, at each of these runs, the electricity prices $\bp$, the disconnection times $\tau_n$, the temporal flexibilities $\alpha_n$, the initial SoCs $\mathbf{s}_n[0]$ and the desired SoCs $s^d_n$ are uniformly sampled from the following datasets (see Figure~\ref{fig:data}).
The daily electricity prices $\bp$ are taken from the day-ahead market prices set by the New England ISO between Nov.\ 2023 and Oct.\ 2024 \cite{iso_new_england_pricing_2024}. We only sampled from the 18 days with the most extreme temperature variations (these days are concentrated in mid-Winter and mid-Summer). Prices were too flat on the other days to support arbitrage. We believe that the days with significant price variations could be representative of a grid that relies more heavily on VREs than New England's (less than 10\% of its electricity is generated by renewables). 
The EVs' desired disconnection times $s^d_n$ come from a dataset collected at a parking lot between Nov.\ 2, 2018 and Jan.\ 2, 2019 \cite{lee_acn-data_2019}. Their initial SoC $\mathbf{s}_n[0]$ and desired SoC $s^d_n$ were collected in 3 charging stations \cite{distributed_electrical_systems_laboratory_desl_epfl_desl-epfllevel-3-ev-charging-dataset_2025}. 
Finally, their temporal inflexibility coefficients $\alpha_n$ were deduced from a survey that asked drivers how much financial compensation they would need to accept a one-hour delay \cite{suel_vehicle--grid_2024}. Following our model of the EVs' cost (Equation \eqref{eq:ev_cost}), we obtain that $\alpha_n$ is equal to the cost of this one-hour delay. As the goal of our experiments is to study the effects of temporal flexibility, we set SoC inflexibility coefficients $\beta_n$ to a prohibitively high value of 10\,\$/kWh$^2$. This amounts to assuming that the drivers attach great importance to receiving their desired charge.

\subsection{The Impact of Battery Wear and Drivers' Temporal Flexibility on Vehicle-to-Grid}
\looseness=-1
We start by conducting experiments to quantify the costs that can be saved by providing V2G services to the grid. An optimal charge schedule obtained on a day with marked price fluctuations is presented in Figure~\ref{fig:soc_battery} and Table \ref{tab:summary_results}. It shows that no arbitrage occurs, as the additional cost from battery wear (0.13\,\$/kWh) exceeds the range of energy prices on that day ($0.22 - 0.13 = 0.09$\,\$/kWh). As a result, an EV cannot make a net profit by exploiting price differences within the day. On the other hand, V2G becomes profitable when the cost of battery wear is reduced by 75\%. A reduction of this magnitude in the future is plausible considering that the price of Li-ion batteries has fallen by approximately 90\% between 2010 and 2023 \cite[p.~55]{irena_world_2024}. In parallel, the average lifespan of a EV batteries continues to increase. Projections suggest that the average number of full cycles of a Li-ion battery in 2030 could be 90\% higher than in 2017 \cite[p.~13]{irena_electricity_2017}. These two trends and the probable accentuation of energy price fluctuations in VRE-intensive grids could make V2G viable in the coming years. 

A second observation that can be made from Figure~\ref{fig:soc_battery} is that profits from V2G are too low to justify delaying EVs' disconnection. Ignoring battery wear for simplicity, an EV can hope to make at most $(0.22-0.13) \times 6.6 \approx \text{\$}0.6$ by delaying its disconnection by one hour. This contrasts with the mean payment of \$31 that drivers would request for this one-hour delay according to our dataset (see Figure~\ref{fig:data}).

\begin{figure}[!t]
\centering
\includegraphics[width=3in]{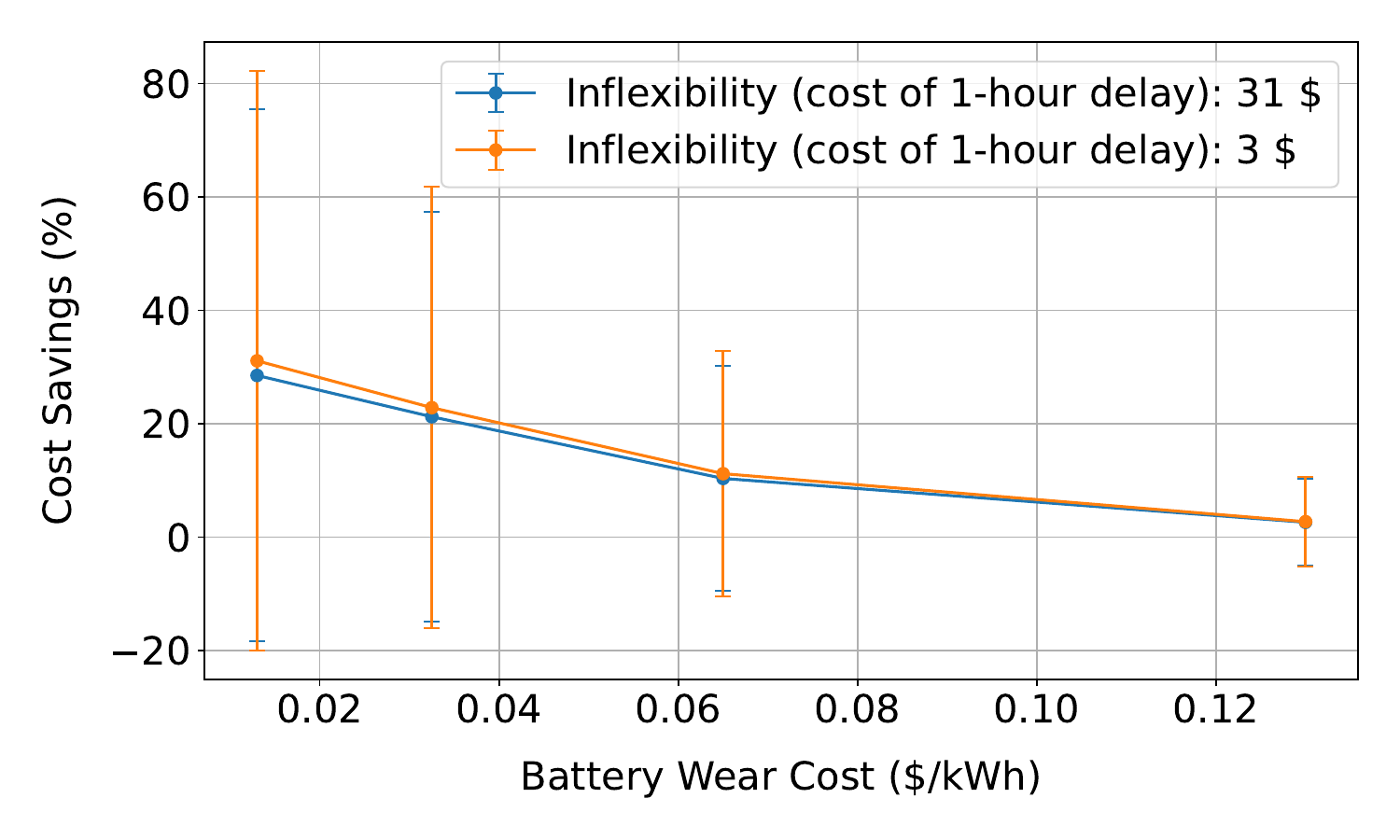}
\caption{Total cost savings of a flexible bidirectional charge schedule with respect to a flexible \textit{unidirectional} charging schedule, as a function of the battery wear cost (mean value and standard deviation over 20 randomly sampled runs, regularly congested station). The lower the battery wear cost, the more savings can be made by allowing bidirectional power flows. Increasing drivers' temporal inflexibility does not increase cost savings because the profits from bidirectional charging are too small to justify delays.}
\label{fig:cost_savings_battery}
\end{figure}

These observations are confirmed over multiple randomly sampled runs, as the costs of battery wear and temporal flexibility are changed (see Figure~\ref{fig:cost_savings_battery}). The lower the battery wear costs, the more savings can be made from V2G. However, dividing the EVs' temporal inflexibility coefficients by 10 is still not enough to make delays profitable.

\begin{figure*}[!t]
\centering
\subfloat[]{
    \includegraphics[width=3in]{soc_regular.pdf}
}
\subfloat[]{
    \includegraphics[width=3in]{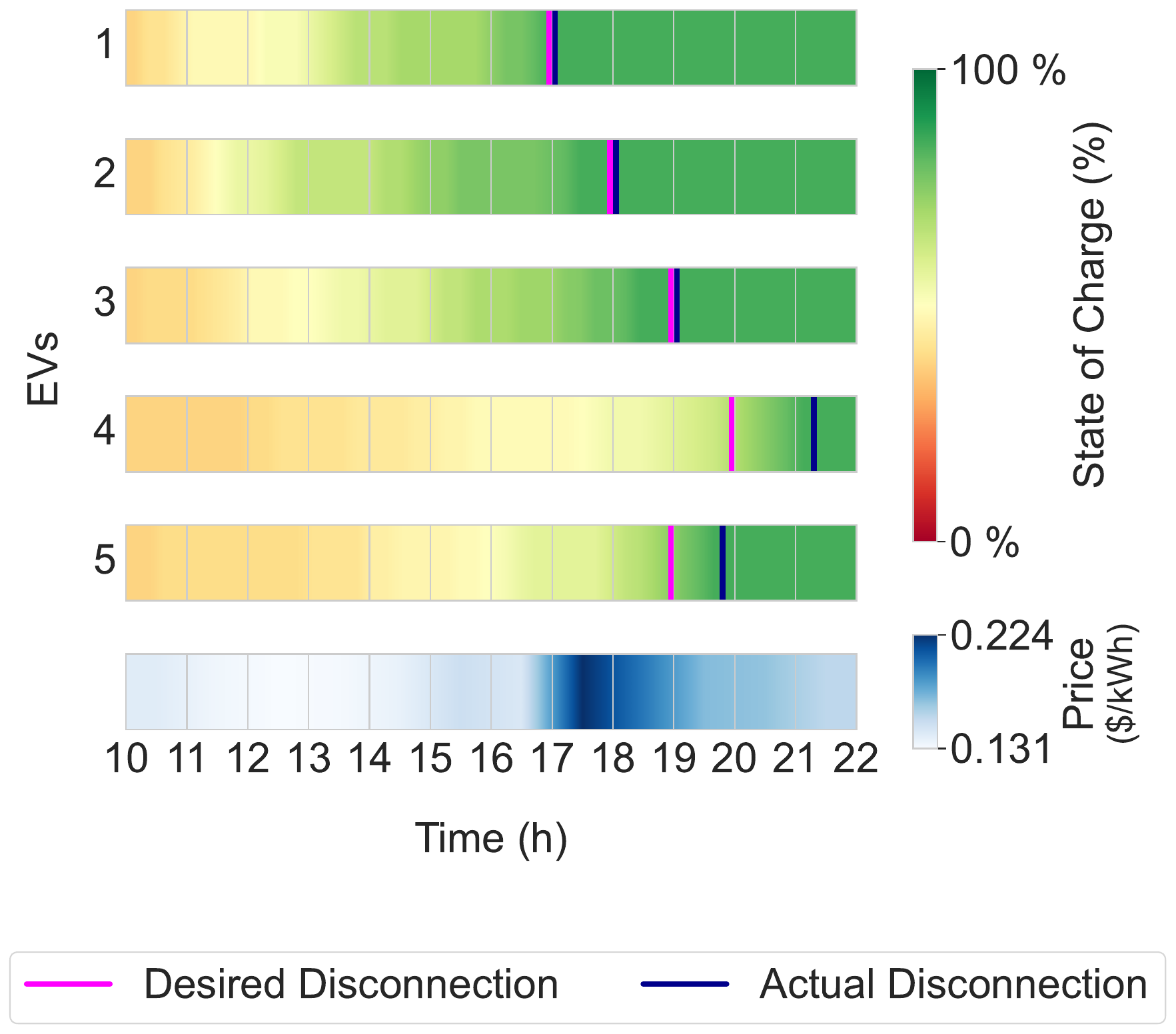}
}
\caption{Optimal flexible scheduling with energy market prices of Jan.\ 17, 2024. (a) Delays are not scheduled with a regular EVCS power limit of 15 kW. (b) Delays become necessary with a lower EVCS power limit of 10 kW.}
\label{fig:soc_congestion}
\end{figure*}

\subsection{The Benefits of Temporal Flexibility for the Alleviation of Station Congestion}
\begin{figure}[!ht]
\centering
\includegraphics[width=3in]{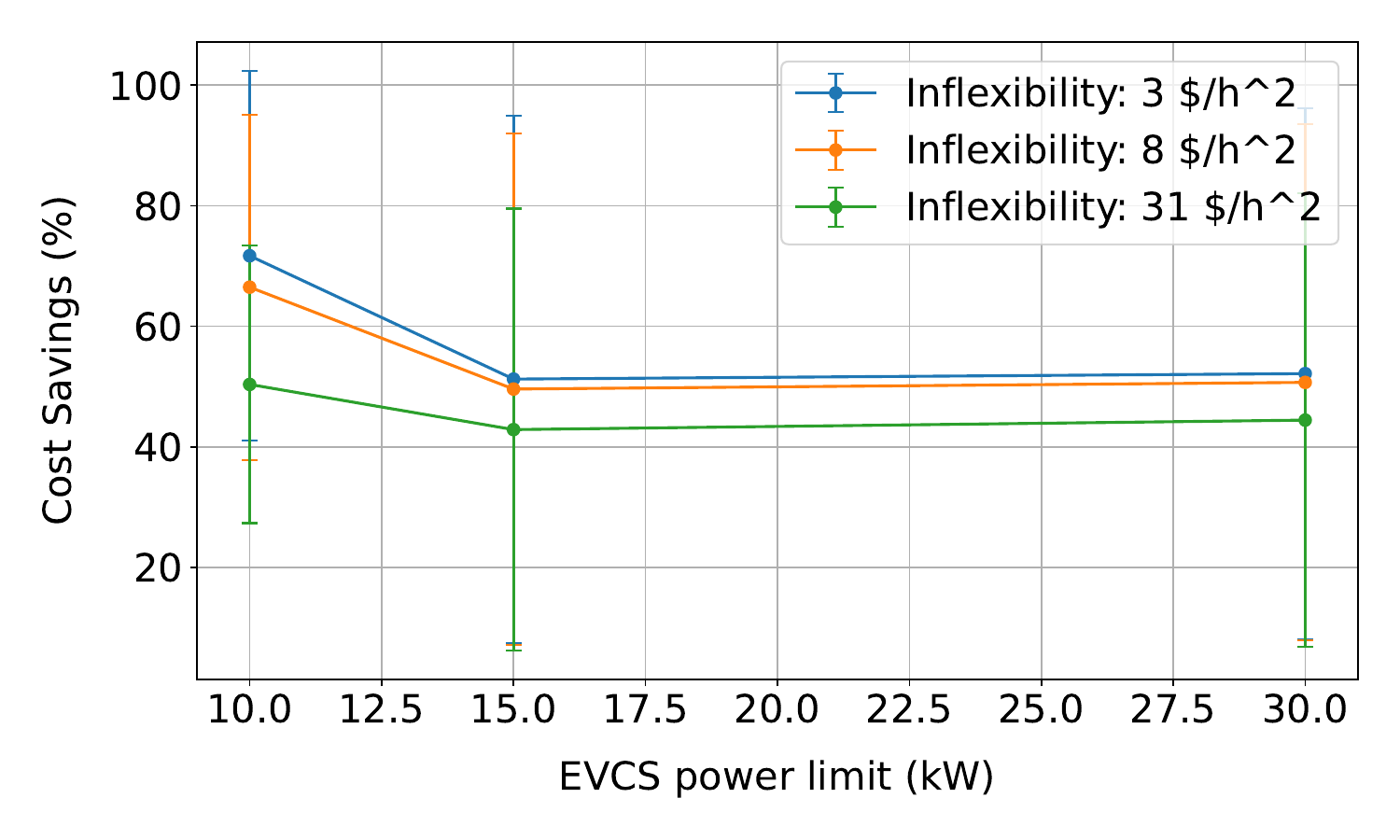}
\caption{Total cost savings of a flexible bidirectional charge schedule with respect to an \textit{inflexible} bidirectional charging schedule, as a function of the charging station's power limit (mean value and standard deviation over 20 randomly sampled runs). 
Below the limit of 15\,kW (which corresponds to an oversubscription ratio of 2.2), cost savings from allowing EV delays seem to increase as congestion worsens. The non-zero cost savings in a non-congested setting (e.g.\ 30\,kW limit) are due to charging requests that are impossible to satisfy because of the the individual EVs' maximum charging rates, rather than the station's capacity.}
\label{fig/cost_savings_flexibility}
\end{figure}

The EVCS starts delaying EVs when it cannot charge them all in time because of high station congestion (see Figure~\ref{fig:soc_congestion} and Table \ref{tab:summary_results}). By setting the charge inflexibility coefficient $\beta_n$ of all EVs to a very high value (10\,\$/kWh$^2$), we have made it preferable for EVs to delay their disconnection rather than endure a partial charge. Consequently, allowing EV delays becomes advantageous as the power limit of the EVCS is reduced. Figure~\ref{fig/cost_savings_flexibility} confirms this observation over multiple randomly sampled runs.

\begin{table}
\caption{Summary of numerical results for Jan.\ 17}
\begin{center}
\label{tab:summary_results}
\begin{tabular}{| c | c | c || c | c |}
\hline
Experiment & Bus & Battery & Average & Average discharged\\
figures  & capacity & wear &  delay & energy for V2G \\
 & (kW) & (\$/kWh) & (min) & (kWh) \\
\hline
\ref{fig:soc_battery}(a),\ref{fig:soc_congestion}(a) & 15 & 0.13 & 0 & 0 \\
\ref{fig:soc_battery}(b) & 15 & 0.03 & 0 & 5.0 \\
\ref{fig:soc_congestion}(b) & 10 & 0.13 & 24 & 0 \\
\hline 
\end{tabular}
\end{center}
\end{table}

\begin{figure}[!t]
\centering
\includegraphics[width=3in]{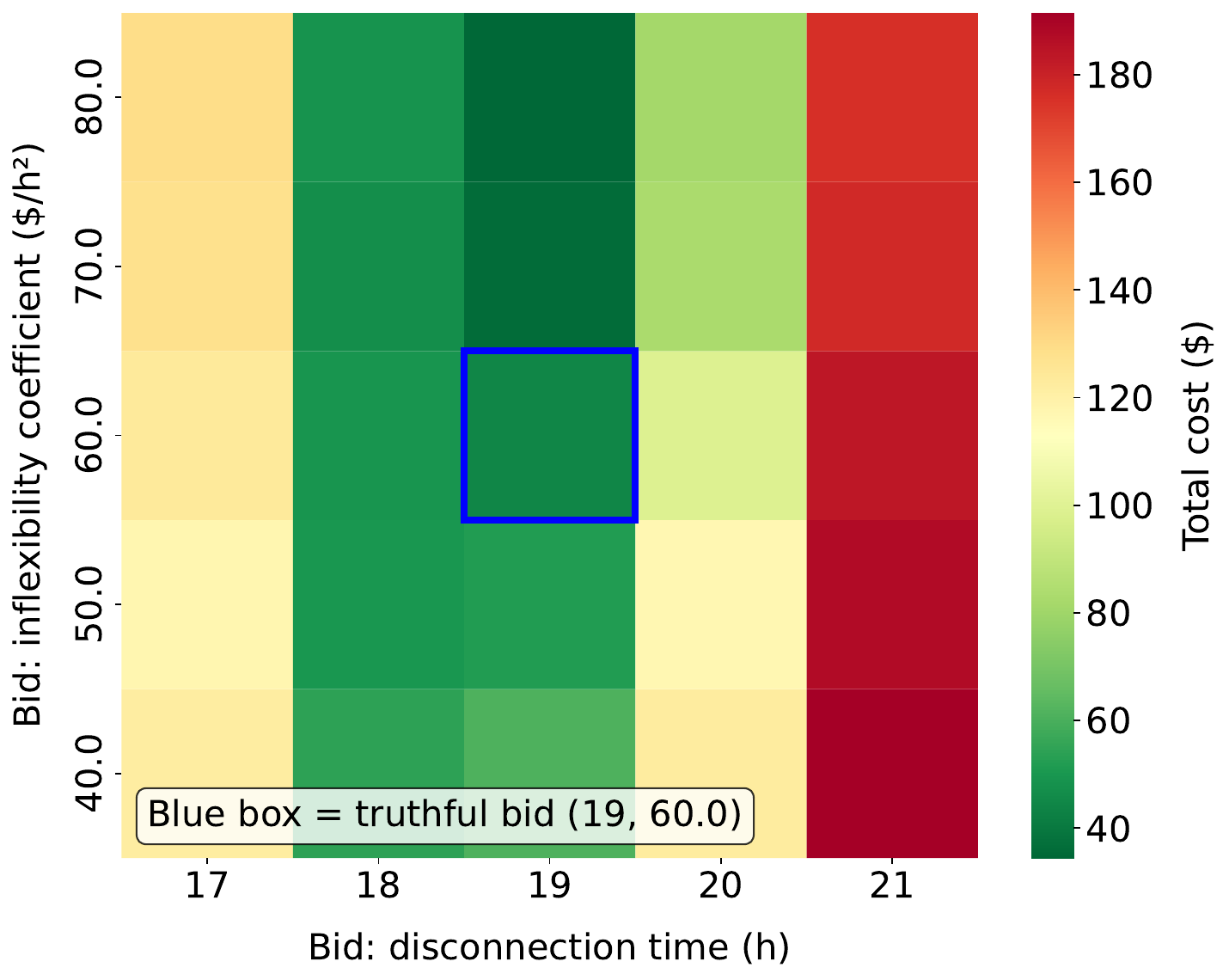}
\caption{Utility of EV 1 as a function of its reported disconnection time and temporal inflexibility in the absence of payments to the station (mean value over 20 randomly sampled runs, highly congested station). Its real charging preferences are indicated with the blue square. The EV benefits from reporting an inflexibility that is higher than its real preference.} 
\label{fig/incentive_lie}
\end{figure}

\subsection{An Empirical Analysis of Vickrey–Clarke–Groves Taxes}
We designed VCG taxes in section \ref{sec:mechanism} to elicit the drivers' private information. In this subsection, we start by demonstrating the added value of this information and the need for its truthful elicitation. We then study the empirical properties of the VCG tax. We conclude with experiments showing the costs saved by an EV driver when participating in the truth-eliciting mechanism.
To study VCG taxes in a setting where misreporting may be beneficial, all experiments in this subsection are run in a highly congested station (EVCS power limit of 10 kW). 

To motivate the need for a truth-eliciting VCG tax, we started by quantifying the added value of drivers' private information. Our experiments focused on the temporal inflexibility coefficients $\alpha_n$. The other preferences are assumed to be known to the EVCS coordinator. We compared the total costs when the coordinator plans a schedule with knowledge of (i) the inflexibility coefficients of each EV, and (ii) only the mean value of these coefficients. Over 20 randomly sampled runs, knowledge of the individual coefficients only resulted in an average 0.15\% reduction in costs. This is not surprising as most of the inflexibility coefficients in our dataset are concentrated in the small range of 30--34\,\$/kWh$^2$ (see Figure~\ref{fig:data}). 
To simulate a scenario with more heterogeneous preferences, we repeated the experiment with a modified dataset. This new dataset was obtained by superposing an up-scaled version of the original dataset (of mean 62\,\$/kWh$^2$) with a dataset of the same size in which all EVs have an inflexibility of 0\,\$/kWh$^2$. In this new experiment, knowledge of the individual preferences yielded a more substantial 33.95\% reduction in costs on average. This result demonstrates that the drivers' private information is of high value to the coordinator when their preferences are heterogeneous.

We designed a truth-eliciting mechanism assuming that strategic drivers have an incentive to misreport their private preferences. Figure~\ref{fig/incentive_lie} confirms that, with reports from the other drivers fixed, an EV has an incentive to report an inflexibility that is higher than their real preference in the absence of incentive scheme.

\begin{figure}[!t]
\centering
\includegraphics[width=3in]{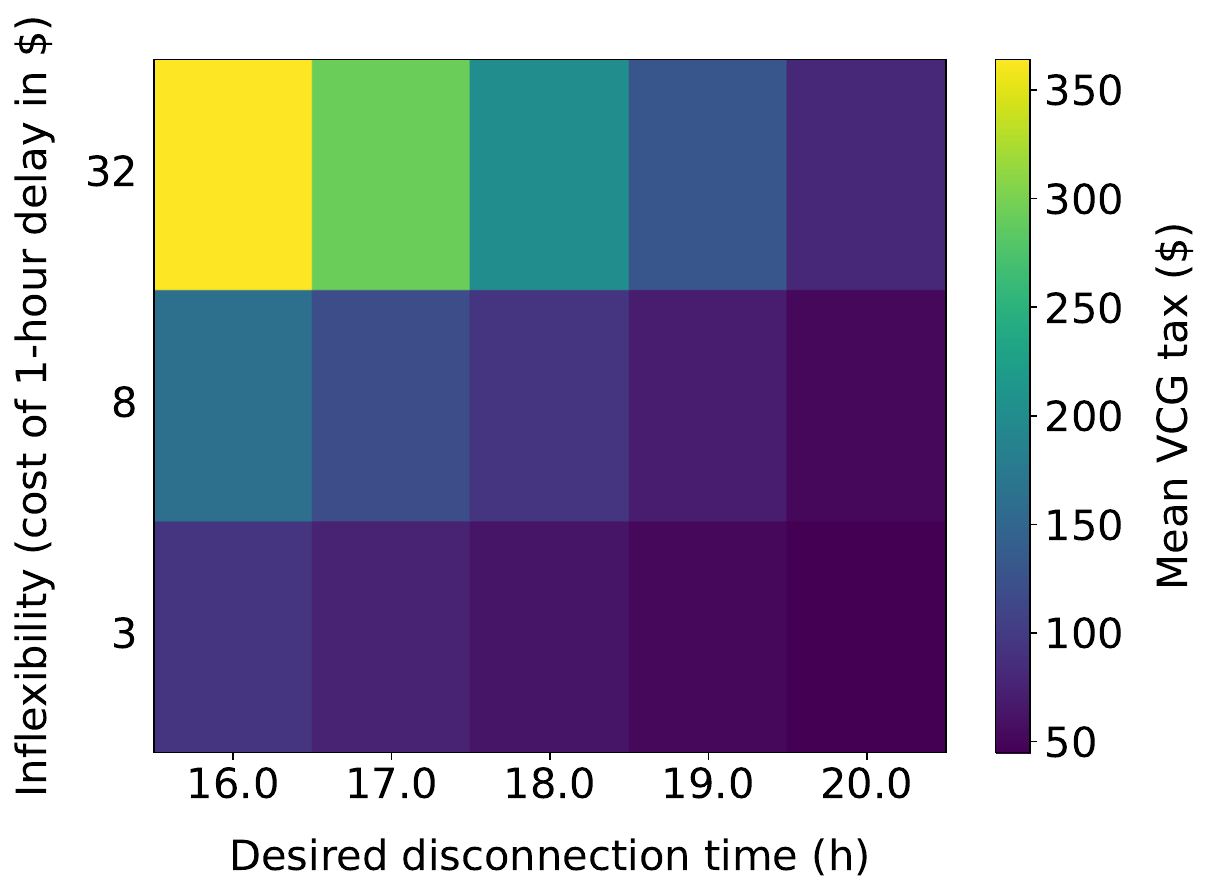}
\caption{VCG tax as a function of the reported charging preferences (mean value over 20 randomly sampled runs, highly congested station.). Earlier desired disconnection times and higher inflexibility lead to higher taxes. The high value of these taxes is partly due to the high oversubscription ratio (which should only result from a temporary malfunction in a well-dimensioned station) and our artificial choice of a high SoC inflexibility of 10\,\$/kWh$^2$.}
\label{fig/vcg}
\end{figure}

\looseness=-1
Having motivated the need for VCG taxes, Figure~\ref{fig/vcg} gives their mean value over multiple runs as a function of a driver's reported disconnection time $\hat\tau_n$ and temporal inflexibility $\hat\alpha_n$.
It shows that VCG taxes tend to increase as a driver requests an earlier disconnection and reports a higher inflexibility. This result matches the interpretation of VCG taxes: a driver pays for the externalities it imposes on its peers. An early inflexible disconnection request in a congested station is likely to be prioritized by the coordinator, to the detriment of the other EVs.

\begin{figure}[!t]
\centering
\includegraphics[width=3in]{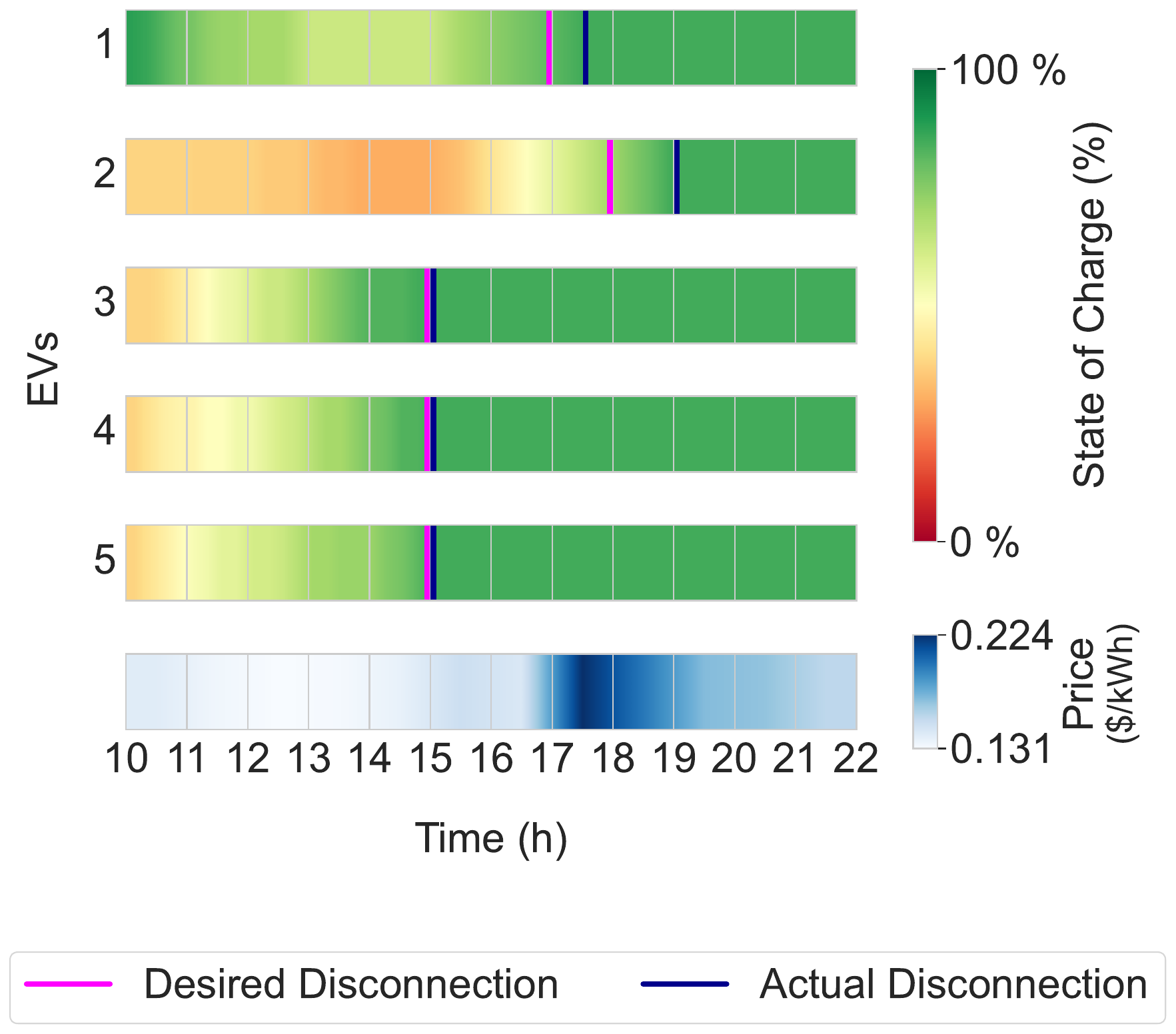}
    \caption{Manually crafted scenario in which the VCG tax is paid by the station to EV 1 ($-\text{\$}104$). The connection of fully charged EV 1 to the station helped to reduce costs for the station by providing V2G services.}
\label{fig/negative_vcg}
\end{figure}

Figure~\ref{fig/negative_vcg} shows a manually crafted example for which a driver receives a net payment from the station. The EV in question connects to the station with a full charge, and its battery is used for V2G (it is technically vehicle-to-vehicle rather than V2G here). As a result, its net externality to the EVCS is positive: it provides a service without itself needing anything from the station. This translates into a payment to the EV by definition of the VCG taxes. However, Figure~\ref{fig/vcg} shows that the average tax is a net payment to the EVCS in practice. Despite the absence of weak budget balance guarantees with VCG, the EVCS makes a net profit. The operation of this mechanism should therefore not require subsidies to the station.

\begin{figure}[!t]
\centering
\includegraphics[width=3in]{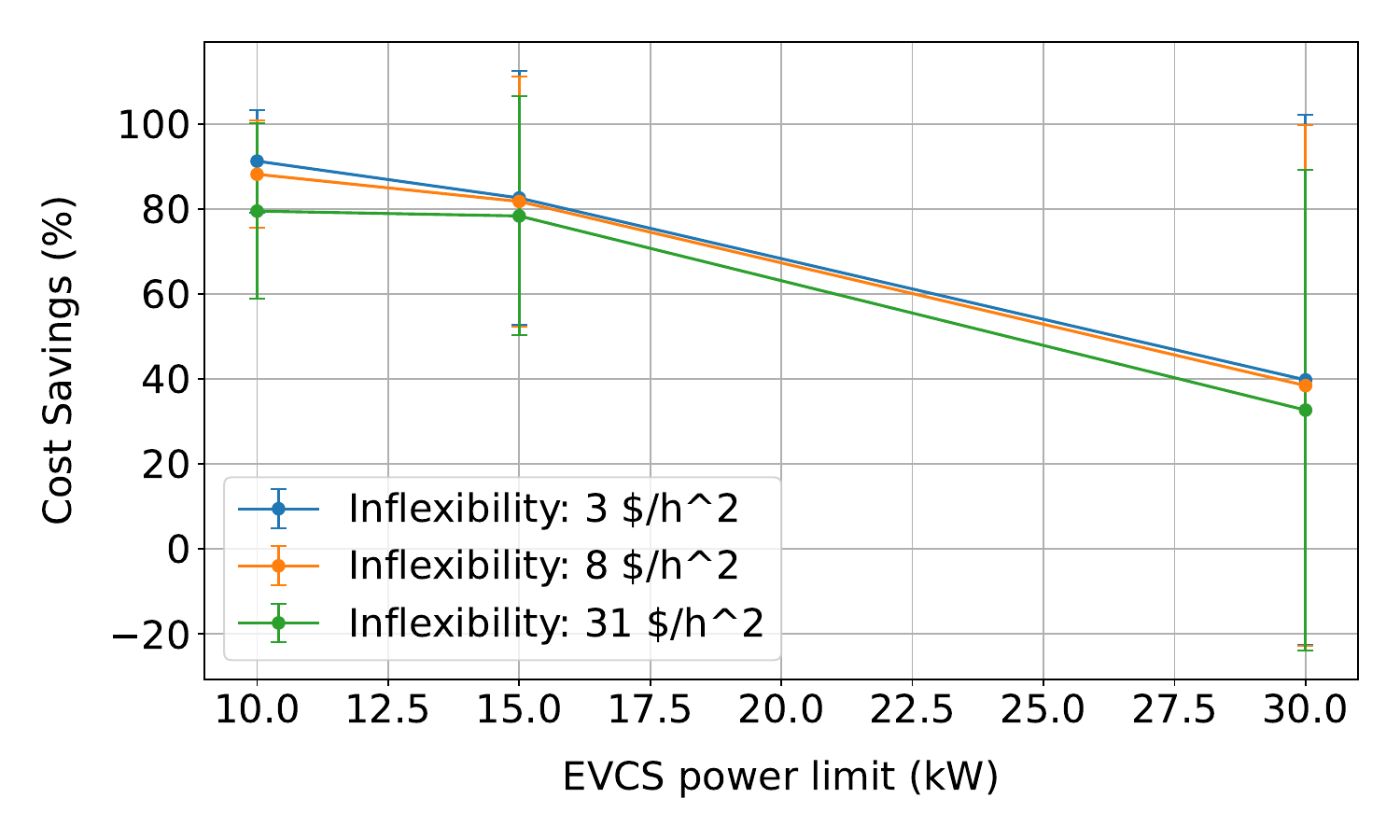}
\caption{Cost savings of an EV from a flexible bidirectional charge schedule (with VCG tax) compared with a a naive charging schedule (where the EVs pay for their energy at market prices), as a function of the station's power limit (mean value and standard deviation over 20 randomly sampled runs). The more congested the station, the more drivers benefit from an optimized schedule, in spite of the truth-eliciting VCG tax.}
\label{fig/cost_savings_vs_linear}
\end{figure}

We showed in Subsection \ref{subsec:vcg} that the VCG taxes are individually rational. An EV always benefits from connecting to the coordinated EVCS rather than remaining discharged. However, drivers might not accept to pay a truth-eliciting tax that could exceed the price of the energy their EV receives. In the absence of correct information on the private preferences, the best alternative for the EVCS coordinator is then to naively charge all EVs in parallel from the start (at the station's maximum capacity). Figure~\ref{fig/cost_savings_vs_linear} shows that it is in the drivers' interest to accept an optimized schedule with VCG taxes instead of this suboptimal outcome. With the naive schedule, intra-day price differences are not exploited and the risk of a partial discharge becomes higher. This risk is exacerbated as the congestion of the station worsens.
In summary, the VCG mechanism elicits the private preferences of the drivers in a way that is acceptable to the individual EV drivers.

\section{Conclusion}
This work addressed the scheduling of temporally flexible and strategic EVs for boosting V2G arbitrage and alleviating station congestion. We formulated the flexible scheduling problem as a mixed-integer quadratic problem and proposed the Alternating Direction Method of Multipliers as a method for approximating its solution. We used the Vickrey–Clarke–Groves mechanism as an incentive scheme to truthfully elicit the drivers' private charging preferences.

Our empirical case study showed that V2G would only become profitable if the cost of EV batteries decreased, if their lifetime increased or if electricity prices fluctuated more. Due to the high cost of drivers' temporal flexibility, the disconnection of EVs is only delayed if the station is too congested to satisfy all charge requests in time. Our analysis also showed that the elicitation of drivers' private information is essential when their preferences are heterogeneous. The truth-eliciting VCG tax can be implemented without subsidizing the charging station.

Future work includes extending the flexible scheduling model to a network of stations. This would add a spatial dimension to the EVs' flexibility, in the spirit of \cite{lv_optimal_2024}. Another important direction would be generalizing the preference elicitation problem to EVs with uncertain charging preferences, similar to what was done in \cite{satchidanandan_efficient_2022}.

\bibliographystyle{IEEEtran}
\bibliography{bibliography}

\vspace{11pt}

\vfill

\end{document}